\documentclass[structabstract]{aa}   
\usepackage{graphicx}
\usepackage{txfonts}
\usepackage{natbib}
\bibpunct{(}{)}{;}{a}{}{,}
\newcommand{\ie}{i.e.}

\newcommand{\lam}{$\lambda$}
\newcommand{\astec}{ASTEC}
\newcommand{\basti}{BASTI}

\newcommand{\cmthree}{cm$^{-3}$}

\newcommand{\kms}{km\,s$^{-1}$}       
\newcommand{\lsun}{L$_{\odot}$}                          
\newcommand{\msun}{M$_{\odot}$}
\newcommand{\rsun}{R$_{\odot}$}

\newcommand{\mearth}{$M_{\oplus}$}
\newcommand{\rearth}{$R_{\oplus}$}
\newcommand{\halpha}{H$\alpha$}                   
\newcommand{\hbeta}{H$\beta$}

\newcommand{\caii}{Ca\,{\sc ii} }

\newcommand{\acenb}{{$\alpha$~Cen~B}} 
\newcommand{\acenab}{{$\alpha$~Cen~A$+$B}} 
\newcommand{\gaia}{{\emph{GAIA}}}
\newcommand{\kepler}{{\emph{Kepler}}}
\newcommand{\corot}{{CoRoT}}			
\newcommand{\corotp}{{CoRoT-7b}}
\newcommand{\corots}{{CoRoT-7}}

\newcommand{\harps}{HARPS}
\newcommand{\uves}{UVES}
\newcommand{\SE}{super Earth}
\newcommand{\SEs}{super Earths}
\newcommand{\teff}{$T_{\rm eff}$}  
\newcommand{\teffs}{$T_{\rm eff}$s}  
\newcommand{\logg}{$\log g$}
\newcommand{\feh}{[Fe/H]}
\newcommand{\vwa}{VWA}
\newcommand{\sme}{SME}
\newcommand{\vald}{VALD}
\newcommand{\starevol}{STAREVOL}
\newcommand{\vsini}{$v \sin i$}
\newcommand{\vrot}{$v_{\rm rot}$}

\newcommand{\vmicro}{$v_{\rm micro}$}
\newcommand{\vmacro}{$v_{\rm macro}$}
\newcommand{\caone}{Ca\,{\sc i}}
\newcommand{\naone}{Na\,{\sc i}}
\newcommand{\naoned}{Na\,{\sc i}\,D}
\newcommand{\mgone}{Mg\,{\sc i}}
\newcommand{\mgoneb}{Mg\,{\sc i}\,b}
\newcommand{\feone}{Fe\,{\sc i}}
\newcommand{\fetwo}{Fe\,{\sc ii}}
\newcommand{\feonetwo}{Fe\,{\sc i}-{\sc ii}}
\newcommand{\milliaa}{m\AA}
\newcommand{\vwaview}{{\tt vwaview}}
\newcommand{\vwaexam}{{\tt vwaexam}}
\newcommand{\vwares}{{\tt vwares}}
\newcommand{\vwatask}{{\tt vwatask}}
\newcommand{\rainbow}{{\tt rainbow}}
\newcommand{\synth}{{\tt synth}}
\newcommand{\maa}{m\AA} 
\begin{document}
   \title{Improved stellar parameters of CoRoT-7\thanks{The CoRoT space mission, launched on December 27, 2006, has been developed and is being operated by CNES, with the contribution of Austria, Belgium, Brazil, ESA, The Research and Scientific Support Department of ESA, Germany and Spain.}}
   \subtitle{A star hosting two \SEs}
   \author{H.~Bruntt\inst{1}
      \and M.~Deleuil\inst{2}
      \and M.~Fridlund\inst{3}
      \and R.~Alonso\inst{4}
      \and F.~Bouchy\inst{5,6}
      \and A.~Hatzes\inst{7}
      \and M.~Mayor\inst{4}
      \and C.~Moutou\inst{2}
      \and D.~Queloz\inst{4}
       }
   \institute{{LESIA, Observatoire de Paris-Meudon, 5 place Jules Janssen, 92195 Meudon, France}
\email{bruntt@phys.au.dk}
           \and{LAM, UMR 6110, CNRS/Universit\'e de Provence, 38 rue F.\ Joliot-Curie, 13388 Marseille, France}
           \and{ESA, ESTEC, SRE-SA, Keplerlaan 1, NL2200AG, Noordwijk, The Netherlands}
           \and{Observatoire de Gen\`eve, Universit\'e de Gen\`eve, 51 Ch.~des Maillettes, 1290 Sauverny, Switzerland}
           \and{Institut d'Astrophysique de Paris, UMR7095 CNRS, Universit\'e Pierre \& Marie Curie, 98bis Bd Arago, 75014 Paris, France}
           \and{Observatoire de Haute-Provence, CNRS/OAMP, 04870 St~Michel l'Observatoire, France}      
           \and{Th\"uringer Landessternwarte Tautenburg, Sternwarte~5, 07778~Tautenburg, Germany}}
	   \date{Received 27 January 2010; accepted 18 May 2010} 
  \abstract
   {Accurate parameters of the host stars of exoplanets are important for 
the interpretation of the new planet systems that continue to emerge.
The \corot\ satellite recently discovered a transiting rocky planet with 
a density similar to the inner planets in our solar system, a so-called \SE.
The mass was determined using ground-based follow-up spectroscopy, 
which also revealed a second, non-transiting \SE.}
   {These planets are orbiting a relatively faint ($m_V=11.7$) G9V star called CoRoT-7.
We wish to refine the determination of the physical properties of the host star,
which are important for the interpretation of the properties of the planet system. 
}
   {We have used high-quality spectra from HARPS@ESO\,3.6m and UVES@VLT\,8.2m.
We use various methods to analyse the spectra using 1D~LTE atmospheric models. 
From the analysis of \feone\ and \fetwo\ lines 
we determine the effective temperature, surface gravity and microturbulence. 
We use the Balmer lines to constrain the effective temperature and
pressure sensitive Mg\,1b and Ca lines to constrain the surface gravity. 
We analyse both single spectra and co-added spectra to identify systematic errors.
We determine the projected rotational velocity and 
macroturbulence by fitting the line shapes of isolated lines.
We finally employ the Wilson-Bappu effect to determine the approximate absolute magnitude.}
{From the analysis of the three best spectra using several methods we find 
$T_{\rm eff}=5250\pm60$\,K,
$\log g = 4.47\pm0.05$, 
$[{\rm M/H}]=+0.12\pm0.06$, and 
$v \sin i = 1.1^{+1.0}_{-0.5}$\,\kms.
The chemical composition of $20$ analysed elements 
are consistent with a uniform scaling by the metallicity $+0.12$~dex.
We compared the $L/M$ ratio with isochrones to constrain the evolutionary status.
Using the age estimate of 1.2--2.3~Gyr based on stellar activity, we determine the mass and radius
$0.91\pm0.03$\,\msun\ and $0.82\pm0.04$\,\rsun. 
With these updated constraints we fitted the \corot\ transit light curve for \corotp.
We revise the planet radius to be slightly smaller, $R = 1.58\pm0.10$~R$_\oplus$, 
and using the planet mass the density becomes slightly higher, $\rho = 7.2\pm1.8 \, {\rm g\,cm}^{-3}$.  }
   {The host star \corots\ is a slowly rotating, metal rich, unevolved type G9V star.
The star is relatively faint and its fundamental parameters
can only be determined through indirect methods.
Our methods rely on detailed spectral analyses that 
in turn depend on the adopted model atmospheres. 
From the analysis of spectra of stars 
with well-known parameters with similar parameters to \corots\ 
(the Sun and \acenb) we demonstrate that our methods 
are robust within the claimed uncertainties. 
Therefore our methods can be reliably used in 
subsequent analyses of similar exoplanet host stars.
}
\maketitle
%
\section{Introduction}
%
\begin{figure*}
\centering
\includegraphics[width=17cm,angle=0]{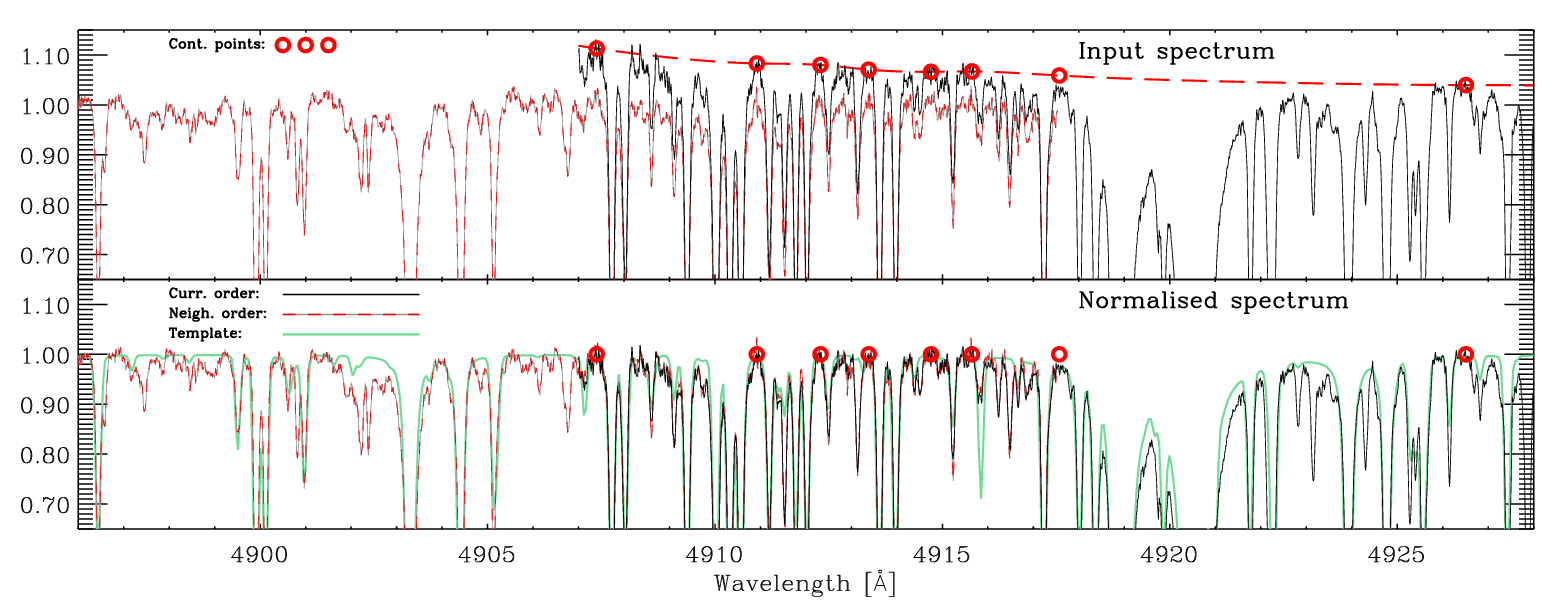} 
\includegraphics[width=17cm,angle=0]{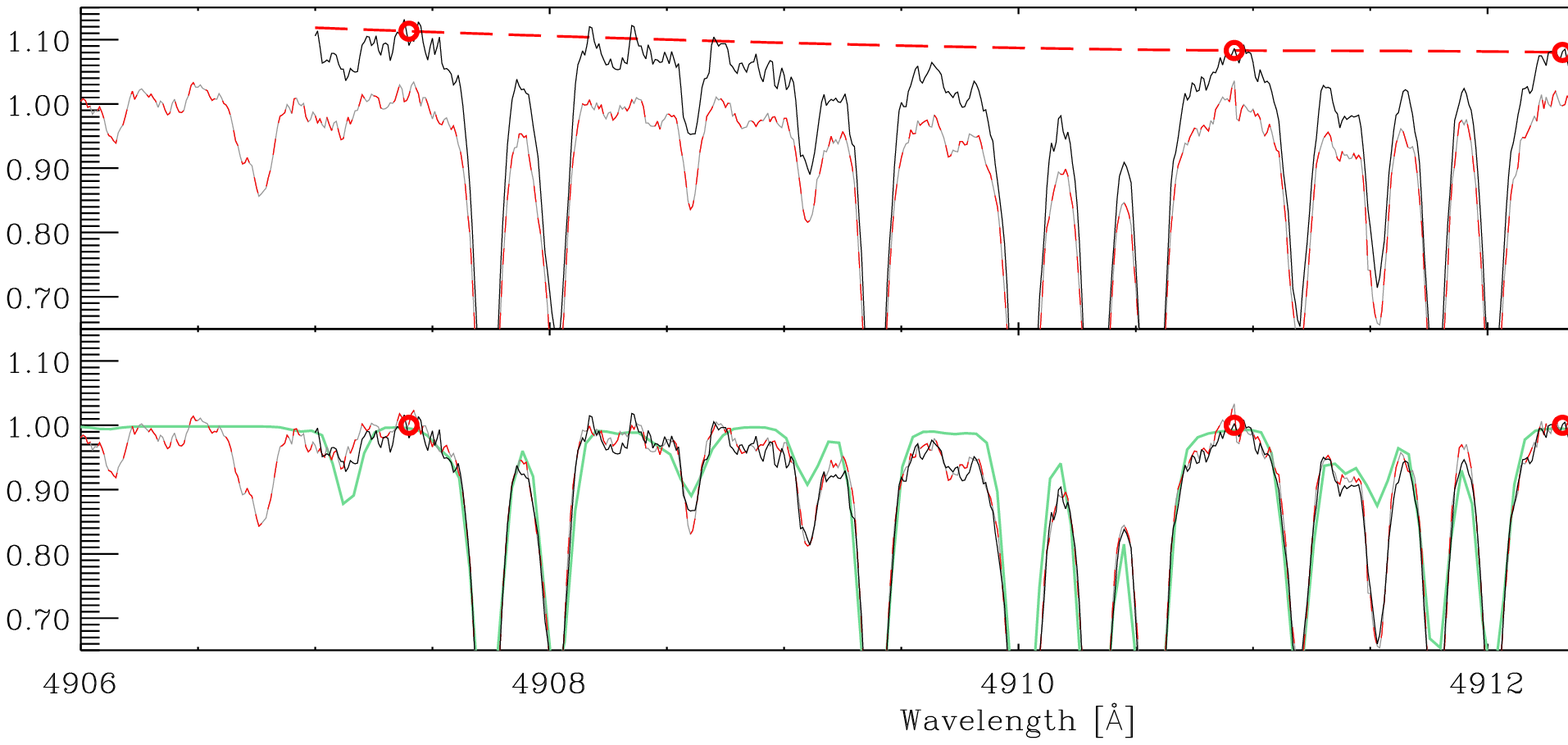} 
\caption{A section of the CoRoT\,7/H1-7 spectrum illustrating how the spectrum is normalised with \rainbow.
The top panels show a wide range and the lower panels show a zoom near the edge of
the same echelle order. The neighbouring order is shown with a short-dashed line. 
The thick long-dashed line is a spline fit to the continuum windows marked by circles. 
The normalised spectrum agrees reasonably well with the template synthetic spectrum (smooth green line). 
The agreement between the two overlapping orders is good and will finally be merged to improve the S/N.
\label{fig:rainbow}}
\end{figure*}
The discovery of the first \SE~planet outside of the Solar 
system with a measured {\em absolute} mass and radius was recently announced,
based on photometric data 
from \corot\ (Convection, Rotation and planetary Transits; \citealt{baglin06}).
This planet, \corotp, has a radius of $1.68\pm0.09$ \rearth\ \citep{leger09}, 
mass $4.8\pm0.8$ \mearth\ \citep{queloz09}, and the orbital period is about 0.85 days \citep{leger09}.
The average density is $5.6\pm1.3$\,g\,\cmthree\ which is similar 
to Mercury, Venus and the Earth \citep{queloz09}. 
Furthermore, a second non-transiting \SE\ has been 
found from radial-velocity monitoring \citep{queloz09}.
These results have only been possible to achieve thanks to 
an extensive ground-based follow-up program of the relatively faint star 
\corots\ (TYCHO~ID 4799-1733-1; $m_V = 11.7$) over more than one year.
    
In the derivation of the planetary parameters, one of the most important factors 
is the correct identification of the host star's fundamental parameters and evolutionary stage. 
It is particularly important to estimate the stellar radius which is 
imperative for determining the absolute planetary radius. A first analysis by several of 
the \corot\ teams has been carried out in \cite{leger09}, based on a spectrum from the
``Ultraviolet and Visual Echelle Spectrograph'' (UVES@VLT\,8.2m) and a preliminary analysis of 
53 co-added spectra from the ``High Accuracy Radial velocity Planet Searcher'' (HARPS@ESO\,3.6m).


Since then, a total of 107 spectra from \harps\ have become available \citep{queloz09}.
These spectra have higher spectral resolution and better signal-to-noise (S/N) 
than the UVES spectrum analysed by \cite{leger09}.
We can therefore now refine the analysis of \corots\ and 
possibly impose stronger constraints on the properties of the system. 
The methods we employ have been developed during the analysis 
of other \corot\ targets \citep{deleuil08, moutou08, rauer09, fridlund10, bruntt09}.
In the current paper we have expanded these tools and we 
will describe our approach in greater detail than previously done.
These tools will be the standard methods to be applied 
for the characterization of future \corot\ targets.

\section{Spectroscopic observations
\label{sec:obs}}

We initially acquired one UVES spectrum which 
confirmed that the star is a dwarf star,
meaning the absolute radius of the planet must be small \citep{leger09}. 
To constrain the mass of \corotp\ a series of 107 spectra were 
collected with the \harps\ spectrograph between March 2008 and February 2009 \citep{queloz09}.

The \harps\ spectrograph has a spectral resolution of $115\,000$ \citep{mayor03},
covering the optical range from 3827~\AA\ to 6911~\AA.
With exposure times of 1800 or 2700\,s, the signal-to-noise 
ratio of the individual spectra varies from $\simeq30$--$90$, 
depending on the conditions at the time of observation. 

We used the data from the standard \harps\ pipeline,
and each order was divided by the blaze function to get 
an approximately rectified spectrum. 
We shifted each spectrum by the radial velocity to set it
to the heliocentric rest frame, using the  
values from \cite{queloz09}. Each spectrum was rebinned to the 
same wavelength grid with a constant step of $0.01$\,\AA.

We suspected that some of the exposures could be affected by reflected Moonlight. 
While such data can be used for measurement of the radial velocity variation, 
scattered light can potentially affect the relative line depths 
and hence systematically affect the analysis. 
In order to identify such potential systematic errors 
we made different combinations of the spectra as presented in Table~\ref{tab:spec}. 
We selected seven spectra acquired during dark time, 
and with the highest S/N, computed in nearly line-free regions around 5830~\AA. 
The co-addition, order per order, of these 7 spectra gives the  H1--7 combined spectrum. 
We also analysed three individual \harps\ spectra with the highest S/N (H1, H2 and H3). 
We finally co-added all \harps\ spectra using 
weights $w \propto {\rm S/N}$ to get the H1-107 spectrum.


\begin{table}
\begin{center}{
\caption{List of the 10 spectra used for the spectroscopic analysis.
\label{tab:spec}}
\begin{tabular}{lccrrr}
\hline
\hline
Spec.   &  Date      & Time      & \multicolumn{1}{c}{$t_{\rm exp}$}  &   &   \\
ID      &  UT        &  UT       &  \multicolumn{1}{c}{ [s] } & \multicolumn{1}{c}{S/N} & \\ \hline
H1      &2008-12-26                        & 04:56     & 2700 &  95 &  \\ 
H2      &2009-01-15                        & 05:39     & 2700 &  90 &  \\ 
H3      &2009-01-17                        & 01:45     & 2700 &  95 &  \\ 
H1--7   &\multicolumn{1}{l}{Combined spec.}&           &      & 235 &  \\
H1--107 &\multicolumn{1}{l}{Combined spec.}&           &      & 700 &  \\
U1      & 2008-09-13                       & 08:39     & 3600 & 470 &  \\ \hline
                                                                     
Ceres   & 2006-07-16                       & 07:50     & 1800 & 220 &  \\ 
Ganymede& 2007-04-13                       & 09:40     &  300 & 340 &  \\ 
Moon    & 2008-08-09                       & 02:39     &  300 & 400 &  \\ \hline 
                                                              
\acenb  &\multicolumn{1}{l}{Combined spec.}&           &      &1030 \\ 
\hline

\\
\multicolumn{6}{l}{\emph{Notes:} H1 to H3 are individual HARPS spectra, H1--7 is 7 co-}\\
\multicolumn{6}{l}{added spectra, and H1--107 is the weighed sum of 107 spectra. }\\
\multicolumn{6}{l}{U1 is the \uves\ spectrum. Ceres, Ganymede and Moon}\\
\multicolumn{6}{l}{are solar spectra from \harps. \acenb\ is 25 co-added }\\
\multicolumn{6}{l}{\harps\ spectra from 2004-05-15.}\\


\end{tabular}}
\end{center}
\end{table}

A preliminary analysis of the UVES spectrum of \corots\ was described in \cite{leger09}.
This spectrum has a lower resolution ($R=65000$) than \harps. 
We include our updated analysis here for completion.

To calibrate our methods, we analysed three HARPS spectra of the Sun, 
available from the ESO/\harps\ intrumentation 
website\footnote{URL: http://www.eso.org/sci/facilities/lasilla/instruments/harps/inst/\\ monitoring/sun.html}.
The spectra were acquired by observing Ceres, Ganymede and the Moon,
and have S/N around 250, 350 and 450, respectively.
We note that the ``Moon'' solar spectrum was observed in 
the high-efficiency EGGS mode which has resolution $R=80000$.
In addition, we analysed a co-added \harps\ spectrum of \acenb, which has similar parameters to \corots. 
\acenb\ has been studied using direct, model-independent methods (interferometry, binary orbit) 
and therefore its absolute parameters (mass, radius, luminosity and \teff) are known 
with high accuracy \citep{mello08}.
We will use this to evaluate our indirect methods that rely on the validity 
of the spectral analysis using 1D~LTE atmospheres.

\section{Versatile Wavelength Analysis (\vwa)
\label{sec:vwa}}

We used the \vwa\ package \citep{bruntt04, bruntt08, bruntt09}
to analyse the spectra listed in Table~\ref{tab:spec}.
It can perform several tasks from normalisation of the spectrum, selection of 
isolated lines for abundance analysis, iterative fitting of atmospheric parameters,
and determination of the projected rotational velocity (\vsini). 
The basic tools of \vwa\ have been described in previous work \citep{bruntt02} 
and here we shall give a more complete description 
of some additional tools in relation to the results we determine for \corots.

\subsection{Normalisation of the spectra
\label{sec:rainbow}}

In Fig.~\ref{fig:rainbow} we illustrate the principles of the \rainbow\ program that
we used to normalise the spectra. The top panels show a wide wavelength range in a single
order and the bottom panels show a zoom near the edge of the same echelle order.
The user must manually identify continuum points by comparing the observed spectrum with a template,
which is usually a synthetic spectrum calculated with the same approximate
parameters as the star. The top panel in Fig.~\ref{fig:rainbow} shows the spectrum
before normalization where eight continuum points have been identified and marked by circles.
A spline function -- optionally a low-order polynomial -- is 
fitted through these points and shown as the long-dashed line. 
The spectrum from the adjacent echelle order is shown with the short-dashed line.
The lower panel shows the normalised spectrum along with the template spectrum.
The agreement with the adjacent order is very good and there is acceptable agreement
with the template. The overlapping orders are finally merged to improve the S/N by up to 40\%.

When the continuum points have been marked for all orders the normalised spectrum is saved.
When the first spectrum has been normalised the continuum points are re-used for the other spectra. 
We then carefully check the normalisation in each case since 
several readjustments are needed, especially in the blue end of the spectrum.

The high S/N in the spectrum shown in Fig.~\ref{fig:rainbow} would indicate that 
the continuum is determined to better than 0.5\%. This is only true if the adopted
synthetic template spectrum is identical to that of the star, 
\ie\ the atomic line list is complete and the temperature and pressure structure 
of the atmosphere model represents the real star. 
From comparison of the template and normalised spectra in several regions (an example is given in Fig.~\ref{fig:rainbow}),
we estimate that the continuum is correct to about 0.5\%, while discrepancies of 1--2\% may 
occur in regions where the degree of blending is high and
in the region of the wide Balmer lines and the \mgoneb\ lines.

\begin{figure}
\centering
\includegraphics[width=9.1cm,angle=0]{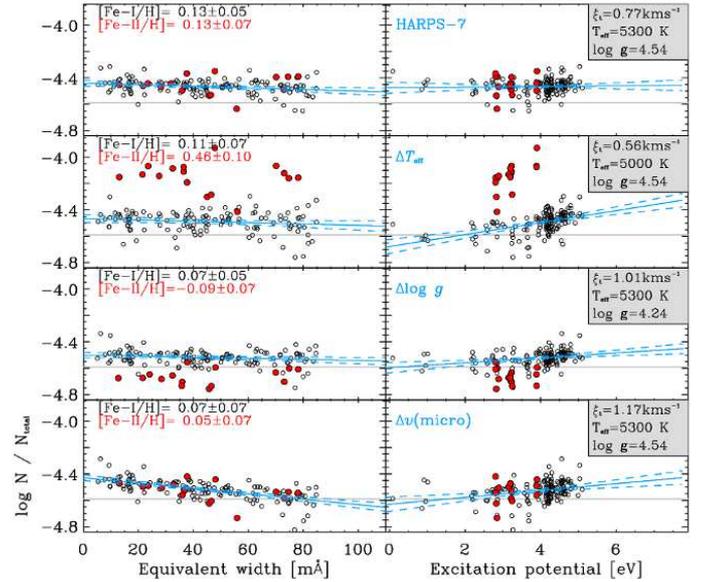} 
\vspace{0.2cm}
\caption{Abundances of \feone\ and \fetwo\ are shown as open and solid red circles, respectively,
and plotted versus equivalent width and excitation potential (plot from the \vwares\ program).
The abundances are from the analysis of the H1-7 spectrum for four different sets of
atmospheric models. The top panel is for the preferred model, the second panel is for
a lower \teff, the third panel for lower \logg, and the bottom panel is for higher \vmicro.
Also indicated is the solar Fe abundances (thin horizontal line) and a linear fit with 95\% confidence
limits indicated by the solid and dashed lines.
\label{fig:vwares}}
\end{figure}
%


\begin{table*}
\begin{center}{\caption{Determined atmospheric parameters for \corots, the Sun and \acenb.}\label{tab:vwa}
\setlength{\tabcolsep}{4.3pt}
\begin{tabular}{lcclc|lllc|cc}
\hline
\hline
       & \multicolumn{4}{c|}{$\langle$\feonetwo$\rangle$}     & $\langle$Mg\,1b$\rangle$  &  $\langle$Ca6122$\rangle$  & $\langle$Ca6162$\rangle$  & $\langle$Ca6439$\rangle$    & \multicolumn{2}{c}{$\langle$Isol.\ lines$\rangle$}  \\ 
Spec.  & \teff\ [K]  & \logg & \multicolumn{1}{c}{\feh} &  \vmicro\ [\kms] & \multicolumn{1}{c}{\logg}  & \multicolumn{1}{c}{\logg}  & \multicolumn{1}{c}{\logg}   & \multicolumn{1}{c|}{\logg}    & \vsini & \vmacro  \\ 

\hline

 H1      & $   5180\pm67$ & $ 4.30\pm0.08$  & $+0.11\pm0.13$ & $ 0.98\pm0.07$ & $3.89\pm0.46 $ &$4.44\pm0.08$ & $4.43\pm0.10 $ &$4.34\pm0.19 $ &  1.3   & 1.2  \\ 
 H1\,\sme& $   5280\pm44$ & $ 4.44\pm0.06$  & $+0.13\pm0.06$ & $ 0.80\pm0.07$ & $4.62        $ &$4.66       $ & $4.53        $ &               &        &      \\ 
 H2      & $   5300\pm41$ & $ 4.46\pm0.08$  & $+0.14\pm0.13$ & $ 0.80\pm0.09$ & $4.02\pm0.47 $ &$4.43\pm0.05$ & $4.56\pm0.06 $ &$4.13\pm0.35 $ &  1.1   & 1.2  \\ 
 H3      & $   5350\pm31$ & $ 4.57\pm0.06$  & $+0.15\pm0.09$ & $ 0.76\pm0.07$ & $4.14\pm0.44 $ &$4.50\pm0.09$ & $4.77\pm0.07 $ &$4.72\pm0.26 $ &  1.1   & 1.2  \\ \hline 
 H1-7    & $   5280\pm35$ & $ 4.48\pm0.06$  & $+0.13\pm0.09$ & $ 0.80\pm0.05$ & $3.94\pm0.52 $ &$4.47\pm0.06$ & $4.60\pm0.08 $ &$4.32\pm0.27 $ &  0.9   & 1.4  \\ 
 H1-107  & $   5300\pm25$ & $ 4.54\pm0.05$  & $+0.13\pm0.08$ & $ 0.77\pm0.05$ & $3.91\pm0.55 $ &$4.51\pm0.06$ & $4.58\pm0.05 $ &$4.45\pm0.18 $ &        &      \\ 
 H1-107\,\sme&$5290\pm44$ & $ 4.49\pm0.06$  & $+0.13\pm0.06$ & $ 0.80\pm0.05$ & $4.43        $ &$4.49       $ & $4.49        $ &               &        &      \\ 

 U1      & $  5300\pm17$ & $ 4.50\pm0.03$ & $+0.11\pm0.06$ & $ 0.70\pm0.08$ & $3.94\pm0.49 $ &$4.42\pm0.05$ & $4.46\pm0.06 $ &$4.41\pm0.19 $ &  $ $   & $ $     \\  \hline 
                                                 
 Ceres   & $  5767\pm17$ & $ 4.44\pm0.03$ & $-0.01\pm0.03$ & $ 1.01\pm0.03$ & $4.50\pm0.08 $ &$4.46\pm0.20$ & $4.43\pm0.10 $ &$4.43\pm0.42$  &  1.4   & 2.1  \\ 
 Ganymede& $  5757\pm17$ & $ 4.43\pm0.04$ & $-0.00\pm0.04$ & $ 0.93\pm0.03$ & $4.51\pm0.10 $ &$4.33\pm0.17$ & $4.38\pm0.08 $ &$4.47\pm0.37$  &  1.1   & 2.3  \\ 
 Moon    & $  5775\pm25$ & $ 4.48\pm0.03$ & $+0.02\pm0.04$ & $ 0.91\pm0.04$ & $4.41\pm0.08 $ &$4.53\pm0.20$ & $4.41\pm0.14 $ &$4.50\pm0.33$  &  2.1   & 2.2  \\ 
\hline
 \acenb  & $  5185\pm25$ & $ 4.50\pm0.03$ & $+0.31\pm0.05$ & $ 0.83\pm0.04$ & $4.01\pm0.50 $ &$4.53\pm0.07$ & $4.51\pm0.05 $ &$4.65\pm0.16$  &  1.0   & 0.8  \\ 
\hline

\\
\multicolumn{11}{l}{\emph{Notes:} Results are from \vwa\ except H1 and H1-107 which are also
given for the \sme\ analysis. The 1-$\sigma$ uncertainties are internal errors.}



\end{tabular}}
\end{center}
\end{table*}
%
%


\subsection{Determination of \teff\ and \logg\ from \feonetwo\ lines
\label{sec:res}}

This part of the \vwa\ program has been described in some detail by \cite{bruntt02}
and we will specify some updated details here. \vwa\ uses 1D~LTE atmosphere models 
interpolated in grids from MARCS \citep{gustafsson08} or modified ATLAS9 models 
\citep{heiter02}. We have adopted MARCS models for this study.
The atomic line data are extracted from \vald\ \citep{kupka99}, 
which is a collection from many different sources. 
The synthetic profiles are computed with \synth\ \citep{vp96}.
The \vwa\ abundances are measured differentially with respect to a solar spectrum.
We have used the FTS spectrum by \cite{kurucz84} which was published in electronic 
form by \cite{hinkle00}. We found that making a differential abundance analysis 
significantly improves the precision on the determined \teff\ and \logg\ (see \citealt{bruntt08}).
We assess the question of accuracy in Sects.~\ref{sec:acenb} and \ref{sec:final}.

\vwa\ consists of three main programs written in IDL. 
Each program has a graphical user interface, called \vwaview, \vwaexam\ and \vwares. 
In \vwaview\ the user can inspect the observed spectrum and select a set of 
isolated lines. They are fitted iteratively by computing synthetic profiles and
adjusting the abundance until the observed and synthetic 
profiles have identical equivalent widths within a fixed wavelength range.
The program \vwaexam\ is used to inspect how well the synthetic profiles
fit the observed lines. The user can manually reject lines or base the rejection
on objective criteria like reduced $\chi^2$ values and the relative line depths. 
It takes about one hour to fit 500 lines on a modern laptop.
This is done using the program \vwatask\ for fixed values of
\teff, \logg\ and \vmicro. The process is then repeated
with various values of these parameters to measure the sensitivity of each line.
The user can then investigate this sensitivity in the program \vwares.

In Fig.~\ref{fig:vwares} we show an example from \vwares\ using the H1--7 spectrum of \corots.
The abundances from \feone\ and \fetwo\ lines are plotted
versus equivalent width (EW; left panels) and excitation potential (EP; right panels) 
for four different sets of atmospheric parameters. 
Open and solid red symbols are used for neutral and ionised Fe lines, respectively.
The top panels are for the preferred
parameters where we have minimised the correlation of \feone\ with both EW and EP
and the mean abundances 
of \feone\ and \fetwo\ agree. The second panel is for \teff\ decreased by 300~K, 
resulting in a clear correlation with EP, and \feone\ and \fetwo\ are
no longer in agreement. For the third panel, \logg\ was decreased by 0.3 dex, leading
to a low mean abundance of \fetwo. The bottom panel is for microturbulence increased by
0.4\,\kms\ which leads to correlations of \feone\ with both EW and EP.
From such analyses we can determine the ``internal'' uncertainty 
on the atmospheric parameters (see \citealt{bruntt08} for a discussion). 

\begin{figure}
\centering
\includegraphics[width=9cm,angle=0]{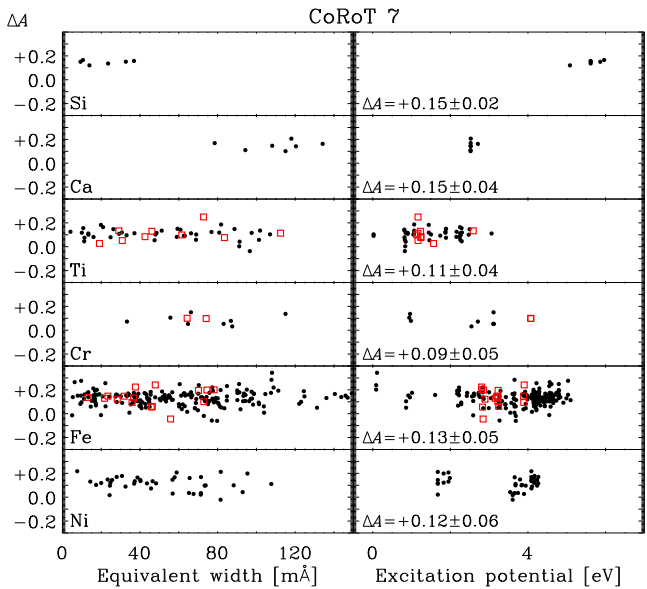} 
\vspace{0.2cm}
\caption{Abundances determined from the H-107 spectrum 
for six elements plotted versus equivalent width and excitation potential.
Solid and open symbols are used for neutral and ionised Fe lines, respectively.
There is no correlation of the abundances and the line parameters,
indicating that the atmospheric model parameters are correct.
\label{fig:corr}}
\end{figure}

In Fig.~\ref{fig:corr} we show an example of the abundances of six elements
determined for the H1--107 spectrum. The mean abundance and rms error is given in the right panels. 
While Fe has the most lines, Ti, Cr and Ni also show no strong correlation 
with equivalent with or excitation potential.


The atmospheric parameters for the six spectra of \corots\ are summarised in Table~\ref{tab:vwa}.
The applied method is indicated in angled brackets in the first row. 
There is good agreement between the results, although the H1 spectrum gives
a systematically lower \teff\ and \logg. This is due to the correlation
between the two parameters as was also noted by \cite{bruntt09}.
They proposed that this degeneracy could be a problem for spectra with relatively 
low S/N (H1 has ${\rm S/N}=57$).
For our final value of \teff\ and \logg\ of \corots\ we adopt the weighted 
mean value of the analysis of the three composite spectra: H1-7, H1-107 and U1: 
$T_{\rm eff} = 5297\pm13$\,K, 
$\log g = 4.51\pm0.02$, 
$v_{\rm micro} = 0.77\pm0.03$\,\kms.
The uncertainties stated here are internal errors.
We will assess the question of ``realistic uncertainties'' in Sect.~\ref{sec:evol}.

\begin{figure}
\centering
\includegraphics[width=9cm,angle=0]{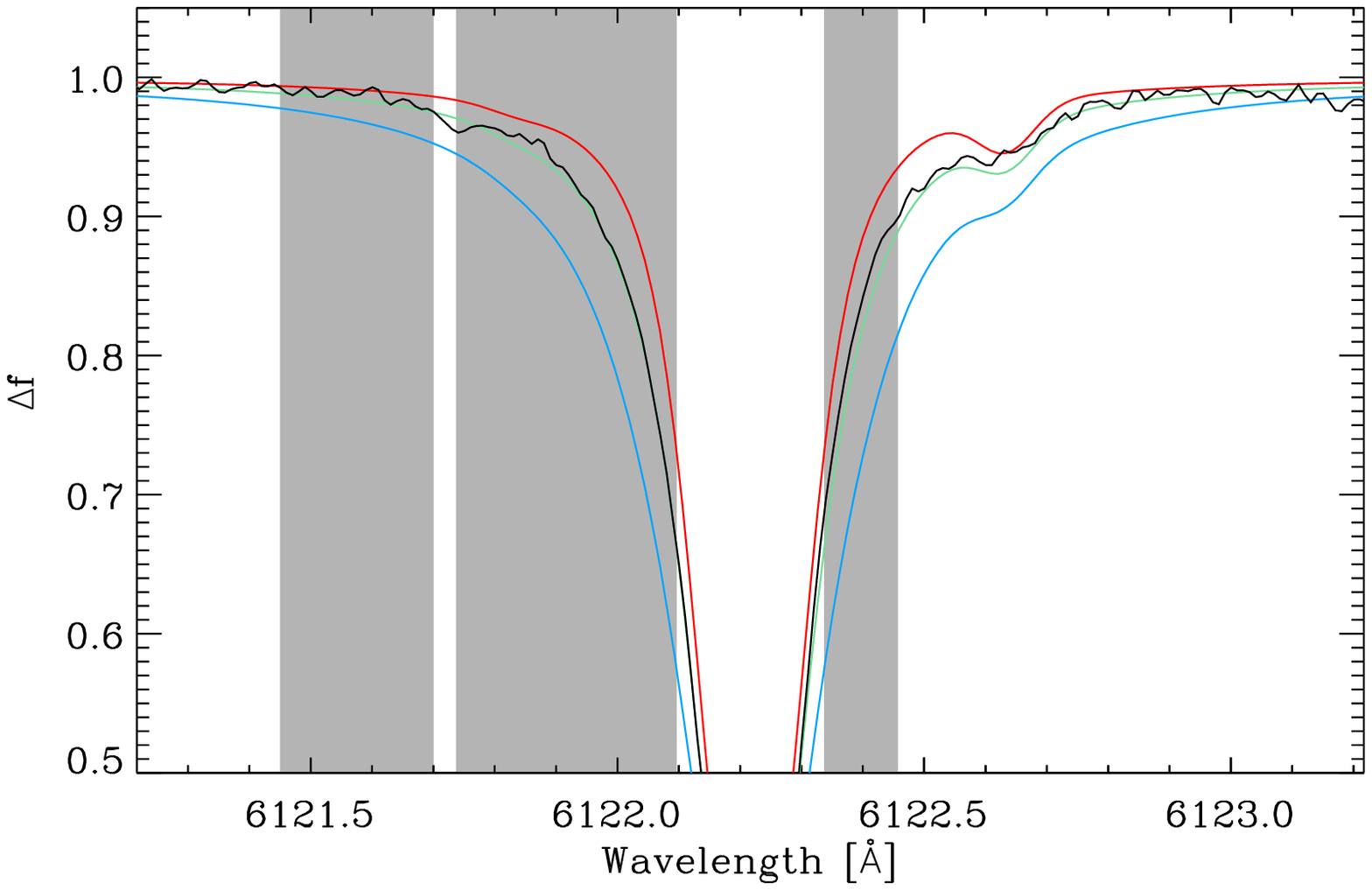} 
\includegraphics[width=9cm,angle=0]{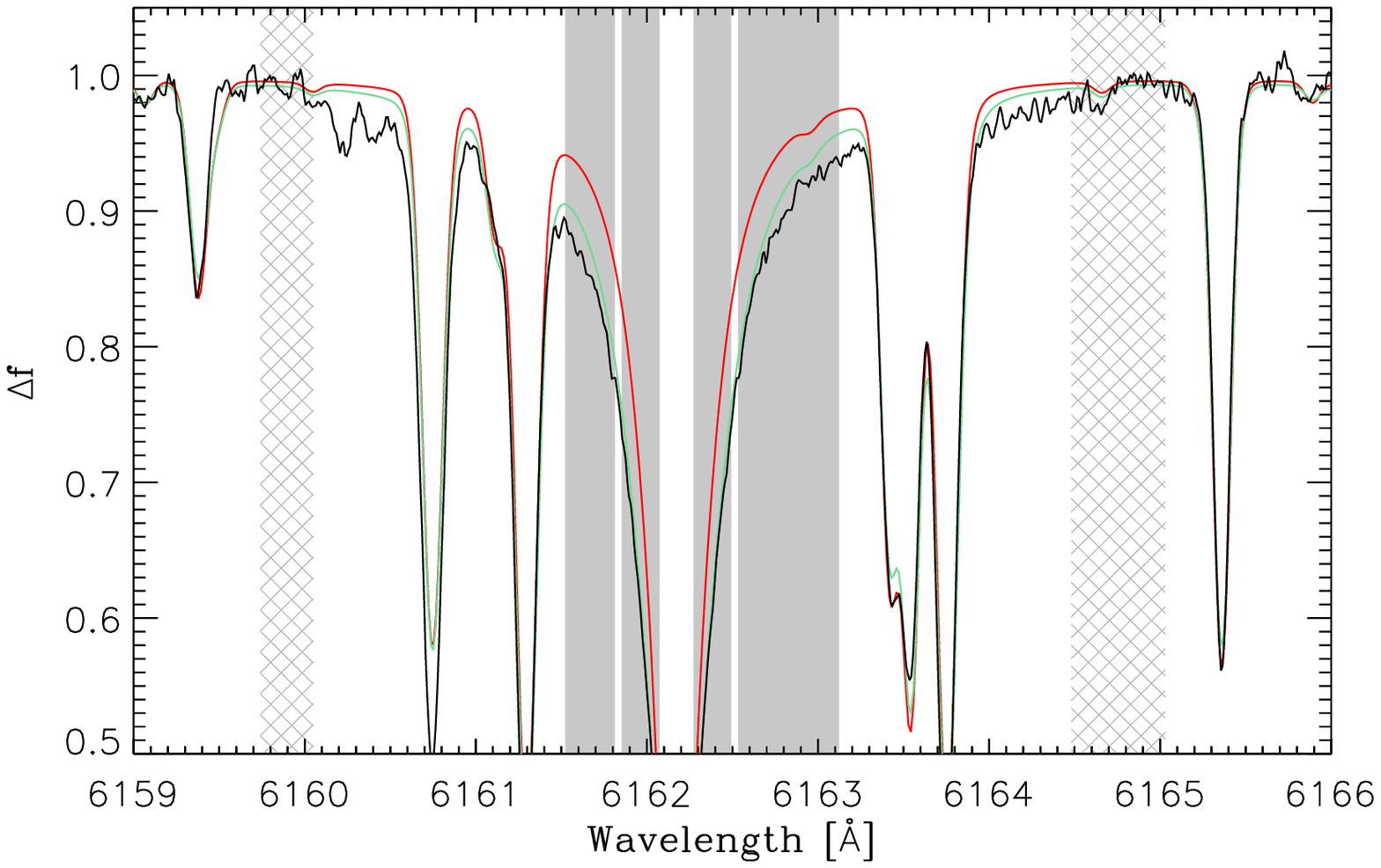} 
\caption{The sensitivity of synthetic Ca lines fitted to the observed spectra 
of the Sun (top panel; Ganymede spectrum) and \corots\ (bottom panel; H1-7 spectrum). 
The synthetic profiles computed for the Sun have $\log g=4.04, 4.44$ and $4.84$.
For the \corots\ spectrum synthetic profiles have $\log g=4.08$ and $4.48$.
In each case a higher \logg\ means the synthetic line becomes wider.
The rectangles mark the areas used to compute the reduced $\chi^2$ and the hatched regions
are used to normalise the spectrum. 
\label{fig:ca}}
\end{figure}

\begin{table}
 \centering
 \caption{The adjusted Van der Waals constants compared to the values from VALD.
 \label{tab:vdw}}
\begin{tabular}{lccc}
     &                 & \multicolumn{2}{c}{$\log \gamma$ [rad\,${\rm cm}^3$/s]} \\ 
Line & $\lambda$ [\AA] & Adjusted & VALD \\ \hline
\mgoneb & 5167.321 & $-7.42$ & $-7.267$ \\
        & 5172.684 & $-7.42$ & $-7.267$ \\
        & 5183.604 & $-7.42$ & $-7.267$ \\ \hline
\naoned & 5889.951 & $-7.85$ & $-7.526$ \\
        & 5895.924 & $-7.85$ & $-7.527$ \\ \hline
\caone  & 6122.217 & $-7.27$ & $-7.189$ \\
        & 6162.173 & $-7.27$ & $-7.189$ \\ \hline
\caone  & 6439.075 & $-7.84$ & $-7.569$ \\

\hline
\hline
 
\end{tabular}
\end{table}

\subsection{Determination of \logg\ from wide lines
\label{sec:ca}}

The surface gravity of late-type stars can be determined from the \mgoneb\ lines,
the \naoned\ and the Ca lines at $\lambda6122$, $\lambda6162$ and $\lambda6439$\,\AA. 
For \mgoneb\ we used only the line at \lam5184\,\AA\ because 
the two lines around \lam5170\,\AA\ are too blended.
We followed the approach of \cite{fuhrmann97} 
to adjust the van~der~Waals constants (pressure broadening due to Hydrogen collisions)
by requiring that our reference spectrum of the Sun \citep{hinkle00} produces 
the solar value $\log g = 4.437$.
In Table~\ref{tab:vdw} we list the adjusted van~der~Waals parameters along with
the values extracted from \vald\ (from \citealt{barklem00}).
Following the convention of \vald\ it is expressed as the logarithm (base 10) 
of the full-width half-maximum per perturber number density at 10\,000~K.
The abundances of the fitted lines are determined from weak lines with ${\rm EW} < 120$\,\milliaa.
The broadening due to \vsini\ and \vmacro\ is determined as described in Sect.~\ref{sec:vsini}.
Examples of fitting the $\lambda6122$\,\AA\ and $\lambda6162$\,\AA\ lines 
are shown in Fig.~\ref{fig:ca} for the Sun (top panel) and \corots\ (bottom panel). 
The hatched regions are used to renormalise the spectrum by a linear fit.
The rectangles mark regions where reduced $\chi^2$ values are computed and they
are used to determine the best value of \logg\ and the 1-$\sigma$ uncertainty.

We found that the \mgoneb\ line in \corots\ is 
not very sensitive and gave lower values ($\log g \approx 4.0\pm0.5$)
than the Ca lines ($\log g \approx 4.5\pm0.1$). 
The reason may be the high degree of blending with weaker lines for such a late type star.
Since the higher value of \logg\
is in good agreement with the result using \feonetwo\ we neglect the results for the \mgoneb\ lines.
There is good agreement for the \logg\ from the individual spectra.
For the value of \logg\ we adopt the weighted mean of the three composite spectra: $\log g = 4.50 \pm 0.02$. 
The stated error does not include systematic errors, see Sect.~\ref{sec:acenb} and \ref{sec:final}.

\subsection{Results for the Sun and \mbox{\boldmath $\alpha$}~Cen~B}
\label{sec:acenb}

It is important to validate that the employed spectroscopic methods produce
trustworthy results. We therefore analysed two fundamental stars for which
\teff\ and \logg\ are known with very high accuracy: the Sun and \acenb.
We analysed three single HARPS spectra of the Sun and one co-added spectrum of \acenb.
The results are summarised in Table~\ref{tab:vwa}.

The parameters from the three solar spectra agree very well with the solar values.
The canonical value for \teff\ is 5777\,K \citep{cox00} and 
\logg\ calculated from the Solar mass and radius is $4.437$.
The largest deviation is 20~K for \teff\ based on the analysis of \feonetwo\ lines.
The surface gravity is constrained by several methods (\feonetwo, \mgone, Ca lines)
but the largest deviation
from the canonical value is only 0.1 dex.
From Table~\ref{tab:vwa} it is seen that some 
lines are less useful for constraining \logg: 
Ca $\lambda6439$\,\AA\ is the least sensitive line. 
For the weighted average, using the Mg and three Ca lines, 
we find excellent agreement for the three Solar \harps\ spectra: 
$\log g = 4.47\pm0.06$, $4.42\pm 0.06$, and $4.43\pm0.06$. 

For \acenb\ we find $T_{\rm eff}=5185\pm25$\,K, $\log g = 4.50\pm0.03$, and ${\rm [Fe/H]} = +0.31 \pm 0.05$
(the quoted uncertainties do not include systematic errors).
For this nearby binary star, \teff\ and \logg\ can be determined by direct methods, \ie\ methods
are only weakly dependent on models. The angular diameter has been measured by \cite{kervella03}. 
Using the updated parallax from \cite{leeuwen07} we determine 
the radius $R=0.864\pm0.005 \, {\rm R}_\odot$. 
The mass has been determined from the binary orbit by \cite{pourbaix02}: $M=0.934\pm0.006 \, {\rm M}_\odot$.
Coincidentally, this mass is nearly identical to that of \corots\ \citep{leger09}.
Combing the mass and radius ($g \propto M/R\,^2$) gives a very accurate 
value of the surface gravity for \acenb: $\log g = 4.538\pm0.008$. 
This is in very good agreement with our spectroscopic determination.
We note that as for \corots, \mgoneb\ is not useful for constraining \logg. 
The \teff\ can be determined from the angular diameter 
and the bolometric flux: $T_{\rm eff}=5140\pm 56$ \citep{bruntt10}.
This is in excellent agreement with the result from \vwa.
\cite{mello08} listed the results of 14 different analyses of \acenb, 
based on different methods and quality of the data. 
Our value of \teff\ is in good agreement with previous determinations 
but our metallicity is slightly higher than most previous estimates.

To conclude, our analysis of the spectra of the Sun and \acenb\ show that we
can reliably determine \teff\ and \logg. 
Since these two stars bracket \corots\ in terms of spectral type, 
we have confidence that the spectroscopic results are robust and 
do not suffer from significant systematic errors.
We will discuss the uncertainties on the spectroscopic parameters in Sect.~\ref{sec:final}.


%
\begin{figure}
\centering
\includegraphics[height=9.2cm,angle=90]{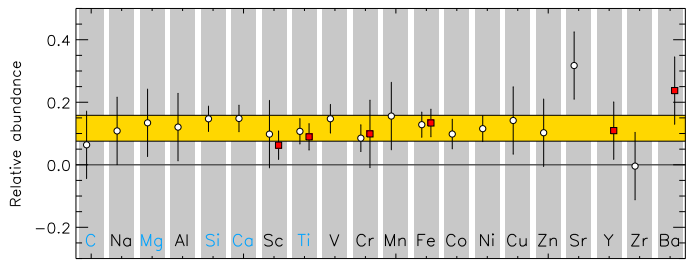} 
\vspace{0.2cm}
\caption{Mean abundances of $20$ elements in \corots\ determined from the H1-107 spectrum.
Circle and box symbols are used for neutral and singly ionised lines, respectively. 
The horizontal bar indicates 
the mean metallicity and the $1$-$\sigma$ error range, $[{\rm M/H}]=0.12\pm0.04$. 
The horizontal line at $0.0$ corresponds to the solar abundance.
\label{fig:chem}}
\end{figure}

\begin{table}
 \centering
 \caption{Abundances relative to the Sun for $20$ elements in \corots. 
Also given is the number of lines used for each element.
 \label{tab:chem}}
 \begin{footnotesize}
\begin{tabular}{llr|llr}

\hline
\hline
  {C  \sc   i} &  $  +0.06        $  &   1   & {Mn \sc   i} &  $  +0.16        $  &   2 \\
  {Na \sc   i} &  $  +0.11        $  &   1   & {Fe \sc   i} &  $  +0.13\pm0.04 $  & 143 \\
  {Mg \sc   i} &  $  +0.13        $  &   1   & {Fe \sc  ii} &  $  +0.13\pm0.04 $  &  16 \\
  {Al \sc   i} &  $  +0.12        $  &   2   & {Co \sc   i} &  $  +0.10\pm0.05 $  &   6 \\
  {Si \sc   i} &  $  +0.15\pm0.04 $  &   6   & {Ni \sc   i} &  $  +0.12\pm0.04 $  &  40 \\
  {Ca \sc   i} &  $  +0.15\pm0.04 $  &   7   & {Cu \sc   i} &  $  +0.14        $  &   1 \\
  {Sc \sc   i} &  $  +0.10        $  &   1   & {Zn \sc   i} &  $  +0.10        $  &   1 \\
  {Sc \sc  ii} &  $  +0.06\pm0.05 $  &   3   & {Sr \sc   i} &  $  +0.32        $  &   1 \\
  {Ti \sc   i} &  $  +0.11\pm0.04 $  &  37   & {Y  \sc  ii} &  $  +0.11\pm0.09 $  &   3 \\
  {Ti \sc  ii} &  $  +0.09\pm0.04 $  &   8   & {Zr \sc   i} &  $  -0.00        $  &   2 \\
  {V  \sc   i} &  $  +0.15\pm0.05 $  &   3   & {Ba \sc  ii} &  $  +0.24        $  &   1 \\
  {Cr \sc   i} &  $  +0.09\pm0.04 $  &   8   &              &                     &     \\
  {Cr \sc  ii} &  $  +0.10\pm0.04 $  &   2   &              &                     &     \\
\hline
\end{tabular}
\end{footnotesize}
\end{table}

\subsection{The chemical composition of \corots
\label{sec:chem}}

The abundance pattern of \corots\ relative to the Sun 
is shown in Fig.~\ref{fig:chem} for the H1-107 spectrum
and in Table~\ref{tab:chem} we list the individual abundances of 20 elements.
We adopted this spectrum since it has the highest S/N but we note that the
other spectra give very similar results.
The mean metallicity is computed from the mean of the metal abundances for species 
with at least 30 lines in the spectrum: Ti, Fe, and Ni. The mean value is
${\rm [M/H]} = +0.12\pm0.04$ where the uncertainty includes the uncertainty
on \teff, \logg\ and \vmicro. 
The horizontal bar in Fig.~\ref{fig:chem} marks the mean value and the 1-$\sigma$
uncertainty range. It can be seen that all elements agree with 
a scaling of $+0.12$~dex relative to the solar abundance. 
For elements with few lines available ($n<3$) we assume an uncertainty of 0.1 dex.

\begin{figure}
\centering
\includegraphics[width=8.8cm,angle=0]{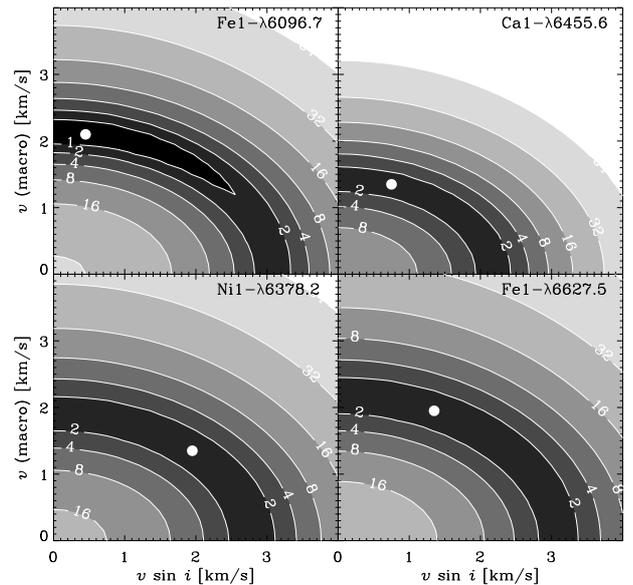} 
\caption{Contours showing the reduced $\chi^2$ values computed for four lines.
The synthetic profiles have been convolved with different combinations of
\vsini\ and \vmacro. The minimum of the surface is marked by a circle.
\label{fig:vsini}}
\end{figure}

\subsection{Determination of \vsini\ and \vmacro
\label{sec:vsini}}

From the detailed profile shapes of isolated lines one can 
ultimately extract information about the granulation 
velocity fields \citep{dravins08}. However, 
this is not possible with our data where each 
single spectrum only has ${\rm S/N} \approx 60$.
The intrinsic shape of a spectral line is determined by several factors \citep{gray08}
but the broadening due to stellar rotation and velocity fields in the atmosphere
can to a good approximation be described by two parameters: \vsini\ and macroturbulence (\vmacro). 
These two parameters describe the projected velocity field due to 
rotation of a limb-darkened sphere
and the movement of granules due to convection, respectively.

To measure \vsini\ and \vmacro\ we selected 64 isolated lines of
different metal species: Ni, Ca, Ti, Cr, and Fe. 
The lines lie in the range 5800--6450\,\AA\ with equivalent 
widths from 25--125\,\milliaa.
For each line we determine the small wavelength shifts needed 
so the observed line core fits the synthetic spectrum. 
This was done by fitting a Gaussian to the line cores of the observed and synthetic spectra.
We then fitted the abundance of the line
for the adopted \teff, \logg\ and \vmicro.
We made a grid of values for \vsini\ and \vmacro\ from 0--6\,\kms\ with steps of 0.15\,\kms.
For each grid point we convolved the synthetic spectrum and computed the reduced $\chi^2$ of the
fit to the observed line. 
In Fig.~\ref{fig:vsini} we show examples of the $\chi^2$ contours for four fitted lines.
The circles mark the minimum of the contour. 
The generally low reduced $\chi^2$ values indicate that our
simple representation of the line broadening is successful. 
It can be seen that there is a strong
correlation between the two parameters. The typical \vmacro\ value for a G9V star is
about 1--2\,\kms\ \citep{gray08}. 
For this range the \vsini\ values for $\chi^2 < 2$ is below 2.5\,\kms\ for nearly all lines.
From this analysis we find mean values of $v \sin i = 1.1^{+1.0}_{-0.5}$\,\kms\
and $v_{\rm macro} = 1.2^{+1.0}_{-0.5}$\,\kms. 
From the analysis of the contours, as shown in Fig.~\ref{fig:vsini},
we can place a firm upper limit of $v \sin i < 3$\,\kms. 



From the transit light curve, \cite{leger09} constrained 
the inclination angle to be $i=80.1\pm0.3^\circ$ (see their Fig.~19). 
Thus, the equatorial rotational velocity is $v_{\rm rot} \approx v \sin i = 1.1^{+1.0}_{-0.5}$\,\kms. 
This result is only valid if we assume that the inclination of the rotation axis of the star 
is the same as the inclination of the orbit.
\cite{leger09} proposed that the rotation period is 
23 days\footnote{We adopt an uncertainty on the rotation period of 2 days.}, 
based on the variation of the \corot\ light curve. 
Using the radius determined in Sect.~\ref{sec:evol} we get \vrot$=1.7\pm0.2$\,\kms,
in agreement with value determined from the spectroscopy.

In Table~\ref{tab:vwa} we list the mean values of \vsini\ and \vmacro\ that 
we have determined for several of the spectra. 
We did not use the U1 spectrum since it has a lower
resolution than the \harps\ spectra. 
We also did not consider the H1-107 spectrum since it is
a combination of so many spectra, which inevitably leads to less well-defined line shapes.

%

\section{Spectroscopy Made Easy (\sme)}


 %
\begin{figure}
\hskip -0.3cm
\includegraphics[width=9.5cm,angle=0]{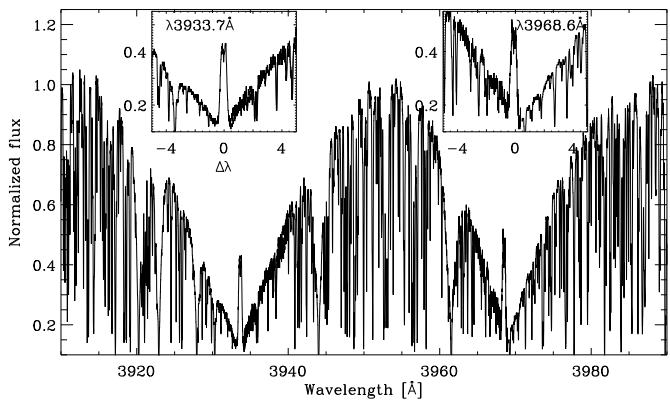} 
\caption{The emission component of the \caii~H\,\&\,K line of \corots. 
The self-reversal in the emission cores is shown in the insets.}
\label{caiiK}
\end{figure}

In an independent analysis of the H1 and H1-107 spectra, 
we used the \sme\ package (version 2.1; \citealt{vp96,vf05}). 
This code uses a grid of stellar models (Kurucz models or MARCS models) 
to iteratively determine the fundamental stellar parameters. This is done by fitting 
the observed spectrum to a synthesised spectrum and minimizing the discrepancies 
through a non-linear least-squares algorithm. \sme\ is based on the philosophy \citep{vp96} that by matching 
synthetic spectra to observed line profiles one can extract the information 
in the observed spectrum and search among stellar and atomic parameters until the best fit is achieved.  

We use a large number of spectral lines, 
e.g.\ the Balmer lines (the extended wings are used to constrain \teff), 
\naoned, \mgoneb\ and \caone\ (for \teff\ and \logg) and metal lines (to constrain the abundances). 
Furthermore, the iterative fitting provides information on micro- and macroturbulence and \vsini.
 
By fitting the extended wings of the \halpha\ and \hbeta\ Balmer lines, 
we determine the \teff\ to be 5200~K and 5100~K respectively. 
Using instead the \naone\ doublet at \lam5890\AA, we find a \teff\ of 5280~K. 
The lower value derived from the \hbeta\ line wings is explained by 
the many metal lines contributing to the profile. 
We tried to use the \mgoneb\ triplet to evaluate \logg\ but as for the \vwa\ analysis 
we found that it is difficult to assign the continuum level, so instead we used the wide 
\caone\ lines. From the \sme\ analysis we find the \logg\ to be 4.43 from \mgone\ and 4.49 from \caone. 
Our evaluation of the metallicity gives ${\rm [M/H]} = +0.13$ and $v_{\rm micro} = 0.80$\,\kms.

The uncertainties using \sme, as found by \cite{vf05}, and based on a large sample  
(more than 1000 stars) are 44~K in \teff, 0.06 dex in \logg\ and 0.03 dex in [M/H],
which we adopt for our \sme\ analysis of \corots.
In a few cases, \cite{vf05} found offsets of up to 0.3 dex for \logg.
When we compare the results for \corots\ for different lines and methods used to constrain \logg, 
we find a scatter of 0.06 dex. 
This is consistent with the results of \cite{vf05} and we assign this as the $1$-$\sigma$ uncertainty. 

In summary, the parameters determined with \sme\ for the H1 and H1-107 spectra
of \corots\ give fully consistent results with the more extensive analysis with \vwa.
Our results from the \sme\ analysis are given in Table~\ref{tab:vwa}.


\section{Absolute magnitude from the Wilson-Bappu effect}
%

 The width of the emission peaks seen in the core of the \caii~H\,\&\,K lines (3934.8 and 3969.7~\AA) in 
late-type stars are directly correlated to the value of \logg, and thus to the mass and radius of the star. 
This implies that the width can be calibrated in terms of the absolute luminosity \citep{gray08}. 
The calibration of the absolute magnitude is of the form:  
$M_V = a\,\log W_0 + b$, where $W_0$ is the width at the zero-level of the emission component, 
and where also the constants $a$ and $b$ need to be properly calibrated. 
This is usually done using data from clusters, and we have used the recent calibration 
of \cite{pace03} who found $a = -18.0$ and $b = 33.2$, with a quoted uncertainty of 0.6~mag on $M_V$. 
 
In Fig.~\ref{caiiK}, we show the \caii\,H\,\&\,K lines of \corots. 
The emission components with self-reversal in the line cores are clearly seen. 
By measuring the width of both the H- and the K-line, following the method of \cite{pace03}, 
we find an absolute magnitude of $M_V = 5.4\pm0.6$. 
Given the spectroscopic effective temperature, 
the location in the Hertzsprung-Russell diagram indicates 
that \corots\ is a main sequence star with spectral type in the range is G8V -- K0V. 
That the star is not evolved is in good agreement with the \logg\ determination.

\begin{table}
\begin{center}{
\caption{Parameters of \corots. 
\label{tab:final}}
\begin{tabular}{lr@{}lll}
\hline
\hline

Parameter   &  \multicolumn{2}{c}{Value} & Unit  & Method \\ \hline
\teff  & $5250$  & $\pm60$    &       K    & Spectroscopy       \\
\logg  & $4.47$  & $\pm0.05$  &            & Spectroscopy       \\
\feh   & $+0.12$ & $\pm0.06$  &            & Spectroscopy       \\

$L/M$& $0.62$   & $\pm 0.08$       & ${\rm L}_\odot/{\rm M}_\odot$ & Spectr.: $L/M\propto T_{\rm eff}^4/g$  \\
\hline

$M$               & $0.91$ & $\pm0.03$ & \msun & Isochrone/tracks \\
$R$               & $0.82$ & $\pm0.04$ & \rsun & Isochrone/tracks \\ 
$L$               & $0.49$ & $\pm0.07$ & \lsun & Isochrone/tracks \\
\logg             & $4.57$ & $\pm0.05$ &       & Isochrone/tracks \\ \hline

\\
\multicolumn{5}{l}{\emph{Notes:} The mass, luminosity and radius are determined}\\
\multicolumn{5}{l}{from comparison with evolution models and rely on the}\\
\multicolumn{5}{l}{age limit of $A<2.3$\,Gyr from \cite{leger09}.}\\



\end{tabular}}
\end{center}
\end{table}
%

%
\section{Evolutionary status
\label{sec:evol}}
%

We will now evaluate the atmospheric parameters determined above for \corots\ and
compare with evolutionary models to constrain the mass, radius and luminosity.

\subsection{Final atmospheric parameters of \corots}
\label{sec:final}

There is generally good agreement for the determination of \teff\ using 
\vwa\ and \sme. With the \vwa\ method we only used \feonetwo\ lines while
with \sme\ we also used the Balmer lines to constrain \teff. As mentioned,
the quoted uncertainties in Table~\ref{tab:vwa} only include the intrinsic
error of the method, \ie\ by varying the model parameters. However, the
temperature and pressure profile of the atmospheric model may not fully 
represent the actual star. From the analysis of the Sun and \acenb, we found
good agreement for their \teff\ and \logg\ determined from model-independent methods
(see Sect.~\ref{sec:acenb}).
Thus, there appears to be no large systematic errors. 
\cite{bruntt10} analysed a larger sample of stars, comparing
the spectroscopic \teffs\ with those from fundamental methods (as done for \acenb\ here)
and found a systematic offset in \teff\ of $-40\pm20$~K. 
We have included this offset to get the final value
$T_{\rm eff} = 5250 \pm 60$\,K.
We used several pressure sensitive spectral features to constrain \logg\ and
the mean value we adopt is $\log g = 4.47\pm0.05$.
For \teff\ and \logg\ we have added systematic errors on 50~K and 0.05 dex,
based on the discussion by \cite{bruntt10}.
The mean metallicity is found to be $[{\rm M/H}] = +0.12\pm0.06$
where we have increased the uncertainty due to the inclusion of systematic
errors on \teff\ and \logg. These are our final estimates for the parameters of
\corots\ and they are summarised in Table~\ref{tab:final}.

Our new results for the fundamental parameters are in good agreement with \cite{leger09}.
They found $T_{\rm eff} = 5275\pm75$\,K as a mean value of different groups
using different spectroscopic analyses of the \uves\ spectrum.
They also used a calibration using 2MASS infrared photometry, taking into account interstellar reddening, 
yielding $5300\pm70$\,K. They find $\log g = 4.50 \pm 0.10$ 
using the \feonetwo\ equilibrium criterion and the \mgoneb\ and \naoned\ lines, 
which is also in good agreement with our value.
\cite{leger09} found a slightly lower metallicity, $[{\rm M/H}] = +0.03 \pm 0.06$
(our revised value for the same spectrum is $[{\rm M/H}] = +0.11 \pm 0.06$).
In that analysis several strong lines were included, while in this study we only used \feone\ lines with
${\rm EW} < 90$\,\maa. 
For other elements (and \fetwo) we included lines with ${\rm EW} < 140$\,\maa.
This choice was made because the strong lines start to be saturated
and are therefore less sensitive to changes in the atmospheric parameters. 
For comparison 250 \feone\ and 18 \fetwo\ lines
were used by \cite{leger09} while we used only 143 and 16, respectively.
In our analysis we used Fe lines in the wavelength range 4880--6865\,\AA, 
while \cite{leger09} included several lines in the blue region (4515--6865\,\AA).
We note that the current version of \vwa\ does not take into account
molecular lines, which start to become a problem for such a cool star,
especially at short wavelengths.


%
%
\begin{figure}
\centering
\includegraphics[width=9cm,angle=0]{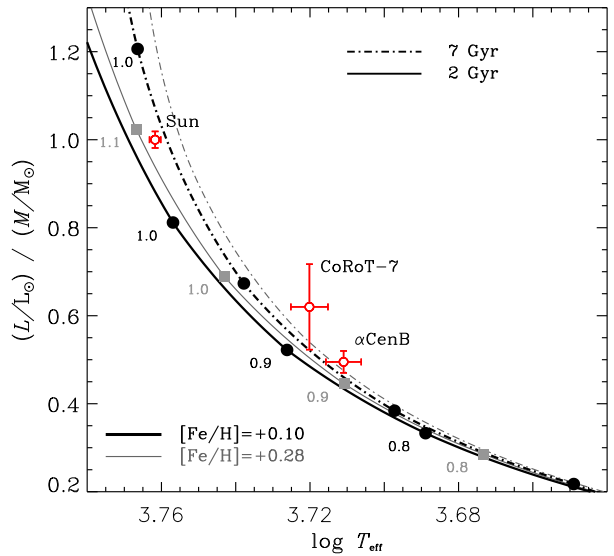} 
\caption{
\basti\ isochrones with different ages and metallicities are shown, 
and filled circles and boxes mark selected mass points.  
The determined $L/M$  ratios for \corots, \acenb, and the Sun are plotted as open symbols.}
\label{fig:hr}
\end{figure}
%

%
%
\begin{figure}
\centering
\includegraphics[width=9cm,angle=0]{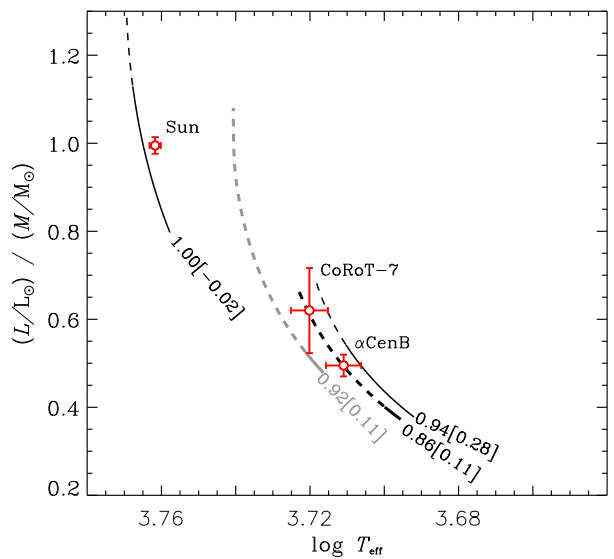} 
\caption{
Four \astec\ evolution tracks are shown for different mass and metallicity,
e.g.\ a track for $1.00\,{\rm M}_\odot$ and ${\rm [Fe/H]} = -0.02$ is shown near the Sun.
The determined $L/M$  ratios for \corots, \acenb, and the Sun are plotted as open symbols.
Dashed lines are used for ages {\em higher} than the adopted limits on the age, 
\ie\ 4.6 Gyr for the Sun, 2.3 Gyr for \corots\, and 6.5~Gyr for \acenb,
while the maximum possible age is 14~Gyr.
\label{fig:jcd}}
\end{figure}

\subsection{Stellar mass, luminosity and radius}


In some cases the modelling of the transit light curve can be 
used to obtain the mean density of the star.
However, as pointed out by \cite{leger09},
the shallow eclipse combined with stellar activity modulating the light curve
seriously hampers such analyses. From the spectroscopic value of \logg\ we
have an estimate of $g = G M/R^2$. Multiplying this with the 
relation $L \propto R^2 T_{\rm eff}^4$ we can eliminate the radius, 
\ie\ $L/M \propto T_{\rm eff}^4 / g$.
Thus, we determine the luminosity-mass ratio: 
$(L/{\rm L}_\odot) / (M/{\rm M}_\odot) = 0.62 \pm 0.08$.
The uncertainty is dominated by the uncertainty on the surface gravity.

In Figs.~\ref{fig:hr} and \ref{fig:jcd} 
we compare this estimate with isochrones from \basti\ \citep{basti04}
and evolution tracks from \astec\ \citep{jcd08}.
These models do not include overshoot 
but this has no impact on low-mass stars such as \corots.
The mixing-length parameter for the \astec\ grid was $\alpha_{\rm ML} = 1.8$.
The models express metallicity in terms of the heavy element mass fraction, $Z$.
To convert each $Z$ to spectroscopic values, we adopted the solar value 
$Z_\odot = 0.0156$ \citep{caffau09} with an assumed uncertainty of $0.002$. 
This corresponds to an increase in the uncertainty of \feh\ by 0.05 dex.

In Fig.~\ref{fig:hr} we show two sets of isochrones with metallicity
${\rm [Fe/H]} = +0.10$ and $+0.28$ for ages of 2 and 7 Gyr, 
with several mass points indicated in the range 0.8 to 1.1 ${\rm M}_\odot$.
The lower metallicity is close to that of \corots\ and the higher metallicity represents \acenb.
The uncertainty on $L/M$ for \corots\ is relatively large, 
so we cannot constrain the mass without further constraints.
Fortunately, \cite{leger09} estimated the age of \corots\ from the rotation period and the 
activity index of the Ca~H\,\&\,K lines: 1.2--2.3 Gyr. 
Adopting this age limit, 
we can estimate the mass and radius from the isochrones:
$M/{\rm M}_\odot = 0.89 \pm 0.03$ and 
$R/{\rm R}_\odot = 0.80 \pm 0.04$.

In Fig.~\ref{fig:jcd} we show four selected \astec\ evolution tracks which
represent the Sun, \corots\ (two tracks), and \acenb.
The dashed part of each track is for ages above these adopted limits:
4.6~Gyr for the Sun ($1.00$\,\msun\ track), 2.3~Gyr for \corots\ (0.92 and 0.86\,\msun),
and 6.5~Gyr for \acenb\ (0.94\,\msun; see \citealt{miglio05} for discussion of the age of \acenab). 
Furthermore, the tracks all end at 14~Gyr. 
It is seen that the Sun is quite well represented, 
although the $L/M$ ratio is quite high at 4.6~Gyr, but 
this is explained by the available track having slightly too low metallicity.
The 0.94\,\msun\ track for \acenb\ agrees with the $L/M$ ratio within the 1-$\sigma$ limit.
For \corots\ the 0.86\,\msun\ track does not reach the determined \teff\ and $L/M$ ratio in
2.3~Gyr. However, for the 0.92\,\msun\ track there is agreement with the \teff\ and $L/M$ ratio.
From similar tracks we determine these limits on the mass and radius of \corots:
$M/{\rm M}_\odot = 0.92 \pm 0.03$ and 
$R/{\rm R}_\odot = 0.83 \pm 0.04$. 
This is in good agreement with the result from the \basti\ tracks.
As our final result, we adopt the mean radius and mass determined from the two sets of models:
$M/{\rm M}_\odot = 0.91 \pm 0.03$ and 
$R/{\rm R}_\odot = 0.82 \pm 0.04$. 

%

\cite{leger09} determined slightly different values for the mass and radius.
They used the \starevol\ evolution tracks (Palacios, private comm.)
with slightly different stellar atmospheric values.
Their values are $M/{\rm M}_\odot = 0.93\pm0.03$ and $R/{\rm R}_\odot = 0.87\pm0.04$ 
(\ie\ $\log g = 4.53\pm0.04$), 
which agree quite well with our revised results given in Table~\ref{tab:final}.

From comparison with the \basti\ and \astec\ models the determined $L/M$ ratio of \corots\ 
seems to be too large, although the uncertainty is large.
In order to determine the luminosity we therefore adjust 
the ratio by $-1\,\sigma$, $(L/{\rm L}_\odot) / (M/{\rm M}_\odot) = 0.54 \pm 0.08$,
and multiply by the inferred mass to get $L/{\rm L}_\odot = 0.49\pm0.07$.
The determined mass and radius from the isochrones correspond to
a surface gravity $\log g = 4.57\pm0.04$, which is slightly higher 
(1.6\,$\sigma$) than the spectroscopic value of $4.47\pm0.05$.






To validate that the \basti\ and \astec\ models can be used for \corots\ 
we also plot the Sun and \acenb\ in Figs.~\ref{fig:hr} and \ref{fig:jcd}. 
For \acenb\ the uncertainty is much smaller than for \corots:
$(L/{\rm L}_\odot) / (M/{\rm M}_\odot) = 0.50 \pm 0.02$.
\cite{miglio05} determined an age of about 6.5~Gyr for the \acenab\ system
and with our metallicity of $+0.3$~dex, there is good agreement with both sets of models.
From comparison with the \basti\ and \astec\ models we get the mass $0.90\pm0.03$\,\msun, 
which agrees well with the dynamical mass of $0.934\pm0.006$\,\msun.
The radius is $0.84\pm0.04$\,\rsun, where the interferometric result is $R=0.864\pm0.005 \, {\rm R}_\odot$. 
Combining the mass and radius from the comparison with the isochrones we
get $\log g = 4.54\pm0.04$, 
which is in good agreement with the spectroscopic value of $4.50\pm0.03$.

\section{Discussion and conclusion}
%

We have presented a detailed spectroscopic analysis
of the planet-hosting star \corots. 
The analysis is based on \harps\ spectra which have
higher signal-to-noise and better resolution than be
\uves\ spectrum used to get a preliminary result \citep{leger09}.
We analysed both individual spectra from different nights 
and co-added spectra and found excellent agreement. 
Only for one of the single \harps\ spectra did we find
a systematic error in \teff\ and \logg, which is explained by the low S/N.

We described in detail the \vwa\ tool which is used
for determination of the atmospheric parameters \teff\ and
\logg\ using \feonetwo\ lines and the pressure sensitive 
\mgoneb\ and Ca lines. We used the \sme\ tool to 
analyse in addition the Balmer and \naone\ lines. We find
excellent agreement between the different methods.

To evaluate the evolutionary status (age) 
and fundamental stellar parameters (mass, radius) we 
compared the observed properties of \corots\ with 
theoretical isochrones. From the spectroscopic \teff\ and \logg\
we can estimate the $L/M$ ratio. We compared this with
isochrones but find that the uncertainty is too large
to constrain the evolutionary status. 
However, by imposing constraints on the stellar age (1.2--2.3~Gyr from \citealt{leger09}) we
can constrain the mass and radius to $0.91\pm0.03$\,\msun\ and $0.82\pm0.04$\,\rsun.
This is a only slight revision of the original value 
from \cite{leger09} who used a lower metallicity. 
The relatively large uncertainty of 7\% on 
the stellar radius directly impacts the accuracy
of the determine radius and density of the transiting planet, \corotp.

We have used the new stellar parameters to fit the transit light curve reported 
by \cite{leger09}. We used the formalism of \cite{gimenez06} with fixed 
limb-darkening coefficients, and we explored the parameter space which is consistent 
with the stellar parameters and their associated uncertainty. The constraints to 
the fit include the orbital inclination ($81.45\pm1.10^\circ$), the phase of transit 
ingress $\theta_1 (0.02785\pm0.00005$), and the ratio of planet-to-star radius 
($0.0176\pm0.0003$). We refer to Sect.~9 in \cite{leger09} for a description 
of the methodology of the fitting procedure.
With the new stellar parameters we determine the radius of 
the planet to be slightly smaller with radius $1.58\pm0.10$~R$_\oplus$ 
(\citealt{leger09} found $1.68\pm0.09$~R$_\oplus$). 
The slightly smaller radius is mainly due to 
our revision of the stellar radius. 

The new stellar mass and the updated inclination were used, 
together with the published values of the ephemeris \citep{leger09} 
and radial velocity semiamplitude \citep{queloz09} to estimate the mass 
of the planet CoRoT-7b as $5.2 \pm0.8 \, {\rm M}_\oplus$. 
Combined with the radius of the planet this results in a density of $7.2 \pm 1.8 {\rm \,g\,cm}^{-3}$.
which is consistent, but slightly more dense than the reported value 
of $5.6\pm1.3$ g\,cm$^{-3}$ in the previous work.


We also analysed spectra of the Sun and \acenb, also
observed with the \harps\ spectrograph. For these stars
the fundamental parameters are known with very good accuracy
and they can therefore be used to validate the methods
we use for the much fainter star \corots. 
We compared the spectroscopically determined 
\teff\ and \logg\ with the values from fundamental methods
for \acenb, \ie\ using the binary dynamical mass and the interferometric
determination of the radius. There is excellent
agreement within 1-$\sigma$, indicating that 
the adopted uncertainties are realistic.
This gives us some confidence that we can use theoretical
evolution models to constrain the radii and masses of stars,
but requires that limits can be put on the stellar age.

The exoplanet host star \corots\ is a slowly rotating, metal rich, type G9V star.
The star is relatively faint and its fundamental parameters
can only be determined through indirect methods. The expected
future discoveries of similar planet systems with \corot\ and \kepler\ will
also be limited by our ability to characterise the host stars.
In the case of \kepler\ we have the additional advantage that for 
the brightest stars the solar-like pulsations can be used to constrain
the stellar radius \citep{jcd10}. This analysis also relies on evolution models but
will be able to constrain the stellar radius to about 2\% \citep{stello09, basu09}.
For most of the \kepler\ targets astrometric parallaxes will be available,
while for \corots\ we must wait for the \gaia\ mission.

\begin{acknowledgements}
We are thankful to Nikolai Piskunov (Uppsala Astronomical Observatory) 
for making \sme\ available to us and for answering numerous questions. 
We are grateful for the availability of the \vald\ database 
for the atomic parameters used in this work.
Based on observations made with ESO Telescopes at the La Silla and
Paranal Observatories under programme IDs 
081.C-0413(C), 082.C-0120, 082.C-0308(A), 282.C-5036(A), and 60.A-9036(A).

\end{acknowledgements}
\bibliographystyle{aa}
\bibliography{bruntt-corot7}

\end{document}